# Quantum Phase Transition from a Spin-liquid State to a Spin-glass State in the Quasi-1D Spin-1 System $Sr_{1-x}Ca_xNi_2V_2O_8$

A. K. Bera and S. M. Yusuf [*]

*Solid State Physics Division, Bhabha Atomic Research Centre, Mumbai 400085, India*

**Abstract:**

We report a quantum phase transition from a spin-liquid state to a spin-glass state in the quasi-one dimensional (1D) spin-1 system $Sr_{1-x}Ca_xNi_2V_2O_8$, induced by a small amount of Ca-substitution at Sr site. The ground state of the parent compound ($x = 0$) is found to be a spin-liquid type with a finite energy gap of 26.6 K between singlet ground state and triplet excited state. Both dc-magnetization and ac-susceptibility studies on the highest Ca-substituted compound ($x = 0.05$) indicate a spin-glass type magnetic ground state. With increasing Ca-concentration, the spin-glass ordering temperature increases from 4.5 K (for the $x = 0.015$ compound) to 6.25 K (for the $x = 0.05$ compound). The observed results are discussed in the light of the earlier experimental reports and the theoretical predictions for a quasi-1D spin-1 system.

[*] Corresponding Author
Fax: +91 22 25505151
E-Mail: smyusuf@barc.gov.in

## I. INTRODUCTION:

Quasi-one dimensional (1D) quantum spin-chain systems have attracted a lot of attentions and are being studied actively due to their various exotic fundamental magnetic properties. In general, according to the Mermin and Wagner theorem, an ideal 1D spin-chain system does not show a long-range (LR) magnetic ordering above $T = 0$ K due to strong quantum spin fluctuations.[1,2] However, for a large number of quasi-1D compounds, LR magnetic ordered ground states, established by a weak inter-chain exchange interaction ($J_\perp$), were found experimentally.[3-6] Besides, Haldane predicted that the ground states and the low lying excitations of a Heisenberg antiferromagnetic (AFM) spin-chain system are totally different for integer and half-integer spins.[7,8] For an integer spin-chain, the ground state is a unique many body singlet, separated by an energy gap (Haldane gap or spin gap) of the order of $2JS^2 \exp(-\pi S)$ from the lowest excited (triplet) state, where $J$ is the coupling between spins within the chain and $S$ is the spin value. On the other hand, a uniform half integer spin-chain system displays a spin-liquid state with a gapless continuum of excitations. Further theoretical study by Sakai and Takahashi predicted a phase diagram in the $D$–$J_1$ plane (where $D$ is the single ion anisotropy and $J_1$ is the ratio of the inter-chain $J_\perp$ over intra-chain $J$ exchange coupling) for the ground state of quasi-1D spin-1 chain systems suggesting a possible quantum phase transition from a spin-liquid state to a Néel-like ordered state.[9] Here the quantum phase transition from the spin-liquid state to the Néel-like AFM ordered state is driven either by the inter-chain exchange interaction $J_\perp$ and/or by the single ion anisotropy. The transition occurs when the strength of the inter-chain exchange interaction *i.e.* $J_\perp$ exceeds and/or the single ion anisotropy $D$ falls behind a certain threshold value. These effects lower the energy of excitations at certain points in the reciprocal space and ultimately induce a transition to a Néel like AFM ordered state. An experimental example of such a Néel like AFM behavior was reported for the quantum spin-1 chain system $CsNiCl_3$.[10-13]

The current experimental focus is to study the quantum phase transition between spin-liquid (non-magnetic) and ordered states by interplaying inter-chain interactions, spin-lattice coupling, spin-vacancies, and/or chemical substitutions. In this respect, the iso-structural 1D spin-chain compounds $A$Ni$_2$V$_2$O$_8$ ($A$ = Pb and Sr), with $S$ = 1 belonging to Heisenberg antiferromagnets have attracted much attention.[14-17] These compounds crystallize in the tetragonal symmetry with a space group of I4$_1$cd.[18] A significant structural feature is that all magnetic Ni$^{2+}$ ions ($S$ = 1) are equivalent with the arrays of edge-shared NiO$_6$ octahedra around the fourfold screw axis and forming screw chains along the $c$-axis (Fig. 1). The screw chains are separated by nonmagnetic VO$_4$ (V$^{5+}$; 3$d^0$, $S$ = 0) tetrahedra and Pb$^{2+}$/Sr$^{2+}$ ions. Studies using different experimental techniques such as magnetization, neutron scattering, nuclear magnetic resonance, and specific heat, have confirmed that both compounds with Pb and Sr show a spin-liquid behavior.[15-17] However, these compounds are located very close to the phase boundary between spin-liquid (non-magnetic) and 3D Néel like ordered states in the Sakai-Takahashi phase diagram.[15-17] Therefore, a small external or internal perturbation can lead to a quantum phase transition from the spin-liquid state to an ordered AFM state in these compounds. Recently He et al.[17] reported a quantum phase transition from the spin-liquid (non-magnetic) state to an "AFM ordered state" in SrNi$_2$V$_2$O$_8$ by a minor substitution of Ca$^{2+}$ ion (with a smaller ionic radius 1.12 Å) at the Sr$^{2+}$ (ionic radius: 1.26 Å) site. He et al.[17] also studied the Ba (larger ionic size) substituted compound Sr$_{0.7}$Ba$_{0.3}$Ni$_2$V$_2$O$_8$. However, no magnetic ordering was found for Sr$_{0.7}$Ba$_{0.3}$Ni$_2$V$_2$O$_8$ compound. It was argued that, upon Ca-substitution, an increase in the strength of the inter-chain interaction ($J_\perp$) occurs due to a reduction of inter-chain distances. This leads to the quantum phase transition from the spin-liquid state (for the parent SrNi$_2$V$_2$O$_8$ compound) to the "AFM state" (for the Ca-substituted Sr$_{0.95}$Ca$_{0.05}$Ni$_2$V$_2$O$_8$ compound) upon achieving the threshold value of $J_\perp$.[17] The inter-chain interactions suppress the energy gap (Haldane gap) between singlet ground state and triplet 1st excited state, and induce an AFM ordering upon reaching the critical value. This is in contrast to the well-established spin vacancies induced magnetic ordering in quantum spin-chain systems,[14, 19-21] where the non-magnetic impurity (Mg or Zn at the Ni-site) breaks the spin-chains and liberates $S$ = 1/2 end chain degrees of freedom resulting a 3D magnetic ordering. For the former case, i.e., the Ca-substituted induced quantum phase transition in SrNi$_2$V$_2$O$_8$, the questions that still remain are, (i) what is the true nature of the magnetic ground (ordered) state of the Ca-substituted compounds? Is it the Néel AFM state or some other state? (ii) is the phase transition from the disordered spin-liquid state to the magnetically ordered state sharp in nature or does it pass through some intermediate magnetic state/states ?, and (iii) what is the critical concentration of Ca-substitution for the onset of the ordered state ? In the present report, we have addressed these important issues by performing a detailed crystal structure and magnetic properties study of the parent compound $SrNi_2V_2O_8$ and a series of Ca-substituted compounds $Sr_{1-x}Ca_xNi_2V_2O_8$ ($x$ = 0.015, 0.025, 0.04, and 0.05). Both X-ray and neutron diffraction techniques were employed to investigate an in-depth crystal structure. Detailed dc-magnetization (both temperature and magnetic field dependent) and ac-susceptibility (temperature dependent, and under different frequencies) measurements were carried out to investigate the true nature of magnetic ground state in the substituted compounds. Our detailed study on the highest Ca-substituted $Sr_{0.95}Ca_{0.05}Ni_2V_2O_8$ compound reveals a spin-glass like magnetic ground state. The observed spin-glass state is in contrast to the reported AFM ground state for the same compound.[17] Nevertheless, our detailed study confirms a quantum phase transition from the spin-liquid state (for $x$ = 0 sample) to a spin-glass like state with Ca-substitution in $SrNi_2V_2O_8$.

## II. EXPERIMENTAL:

Polycrystalline samples of $Sr_{1-x}Ca_xNi_2V_2O_8$ ($x$ = 0, 0.015, 0.025, 0.04, and 0.05) were synthesized by the solid state reaction method. Stoichiometric amounts of high purity (≥ 99.99%) $CaCO_3$, $SrCO_3$, NiO, and $V_2O_5$ were initially mixed using an agate mortar pestle and placed in platinum crucibles. The well ground powders were decarbonated at 950 °C for 36 hours and then pressed into a pellet form. The palletized mixtures were heated for total 72 hours at 1050 °C in air with the reactants being remixed and re-palletized at frequent intervals.

The powder X-ray diffraction (XRD) measurements were performed on all samples at 300 K using a Cu $K_\alpha$ radiation over the scattering angular range $2\theta$ = 10°–90° (the magnitude of the scattering vector $Q$ [= $(4\pi\sin\theta)/\lambda$] range of 0.71–5.77 Å$^{-1}$ ) with an equal $2\theta$ steps of 0.02°.

Neutron diffraction patterns for all samples were recorded at 300 K by using the five linear position sensitive detector (PSD) based powder diffractometer ($\lambda$ = 1.2443 Å) at Dhruva research reactor, Trombay over the scattering angular range of 5°–138° covering a $Q$ range of 0.35–9.4 Å$^{-1}$. The diffraction data were analyzed by the Rietveld method using the FULLPROF program.[22]

The dc-magnetization measurements as a function of temperature ($T$) and magnetic field ($H$) were carried out using a quantum design SQUID magnetometer and a Vibrating Sample Magnetometer (VSM), respectively. The temperature dependent magnetization curves [$M(T)$] for all samples were recorded under 100 Oe field in the warming cycle over the temperature range of $2 \leq T$ (K) $\leq 300$ in zero-field-cooled (ZFC) condition. For $x = 0.015$ and $x = 0.05$ samples, additional magnetization curves have also been recorded in field-cooled (FC) condition under 100 Oe field. For the highest Ca-substituted compound $Sr_{0.95}Ca_{0.05}Ni_2V_2O_8$, additional $M(T)$ curves were recorded under magnetic fields of 50, 80, 200, 500, 1000, and 2000 Oe over the temperature range 2–30 K in zero-field-cooled condition. The $M(H)$ curves over all four quadrants (including the initial leg) were recorded at 2 K for the highest substituted compound $Sr_{0.95}Ca_{0.05}Ni_2V_2O_8$. For the $M(H)$ measurement, the sample was first cooled from room temperature to 2 K under a zero magnetic field.

The ac susceptibility ($\chi_{ac}$) measurements on the $Sr_{0.95}Ca_{0.05}Ni_2V_2O_8$ compound were performed by using a quantum design SQUID magnetometer over the temperature range of 2–20 K and in the frequency range of 1–987 Hz.

## III. RESULTS AND DISCUSSION:

### A. Crystal Structure Study:

The crystal structures of $Sr_{1-x}Ca_xNi_2V_2O_8$ ($x = 0$, 0.015, 0.025, 0.04, and 0.05) compounds were studied from simultaneous Rietveld analysis of both X-ray and neutron powder diffraction patterns recorded at 300 K (Figs. 2 and 3). The refined values of various structural parameters such as, lattice constants, fractional atomic coordinates, and isotropic thermal parameters are given in Table I. All compounds crystallize in the tetragonal symmetry with space group I4$_1$cd as reported earlier for the parent compound.[17] The variation of lattice constants and the unit cell volume as a function of Ca concentration is shown in Fig. 4. With increasing Ca-concentration, a continuous decrease in the lattice constant $a$ has been observed, whereas, $c$ value increases. An overall decrease in the unit cell volume has been found with substitution. The above observations are in consistent with the earlier report by He et al.[17] The $NiO_6$ octahedra and $VO_4$ tetrahedra are found to be slightly distorted. Within a unit cell, oxygen atoms are located at four different crystallographic sites which are labeled as O1, O2, O3, and O4 [Fig. 1(a)]. Along the $c$ axis, $NiO_6$ octahedra are connected by sharing their edges via O2 and O3 oxygen ions, and form the screw chains. Such screw chains are connected by $VO_4$ tetrahedra by sharing corners. Here, two corners of a $VO_4$ tetrahedron are connected to one screw chain via O2 and O3 oxygen ions; one of each from the two adjacent $NiO_6$ octahedra of a given chain. While, other two corners of the $VO_4$ tetrahedron are connected to an adjacent $NiO_6$ screw chain via O1 and O4 oxygen ions; one of each from the two adjacent $NiO_6$ octahedra [Fig. 1(a)]. Therefore, the possible pathways for the intra-chain superexchange interaction along the $c$ axis are Ni-O2-Ni and Ni-O3-Ni [Fig. 1(a)]. For both pathways, the respective bond lengths (Ni-O2 and Ni-O3, respectively) as well as bond angles (Ni-O2-Ni and Ni-O3-Ni, respectively) have almost similar values (Table II and Table III, respectively). On the other hand, two types of inter-chain superexchange interaction pathways between Ni ions are present for these compounds [Fig. 1(a)]. The first type of inter-chain exchange pathway couples two $Ni^{2+}$ ions (of adjacent chains) with same $c$ axis fractional coordinate through the O2-V-O4 and O1-V-O3 bridges [Fig. 1(a)]. The other type of inter-chain exchange pathway couples two $Ni^{2+}$ ions (offset relative to each other by $c/4$ along the $c$ axis) via the O1-V-O2 and O3-V-O4 bridges. For the present system, the intra-chain AFM superexchange interactions $J$ (between two nearest neighbor $Ni^{2+}$ ions along the $c$ axis) dominate due to the shorter pathways along the chain as compared to that between chains. However, the inter-chain exchange interaction $J_\perp$ seems to be not negligible as the screw chains are bridged by $VO_4$ tetrahedra. The strength of $J_\perp$ depends on the exchange propagation pathways between Ni-chains which are effectively the distances between $Ni^{2+}$ ions of the adjacent chains as reported by Mastoraki et al.[23] for a similar type of compound $PbNi_{1.88}Mg_{0.12}V_2O_8$. We will return to the discussion on $J$ and $J_\perp$ later. An important observation here is that the Ni-Ni distances within a given $ab$ plane are not equal along two basal axes and change oppositely with Ca-substitution, [Fig. 4]. With an increasing Ca-concentration, a decrease in the Ni-Ni distances along *b* axis and an increase along *a* axis have been found. The Ni-Ni distance within a given chain along the *c* axis remains unchanged with Ca-substitution. Therefore, an anisotropic inter-chain exchange interaction needs to be considered for these compounds.

**B. dc Magnetization Study:**

The earlier reports on the parent compound $SrNi_2V_2O_8$ established that it has a spin-liquid ground state.[16, 17] Since the present study is based on a series of Ca-substituted compounds $Sr_{1-x}Ca_xNi_2V_2O_8$ ($x$ = 0, 0.015, 0.025, 0.04, and 0.05), we have carried out dc magnetization measurements on the parent compound as well, and reconfirmed its spin-liquid ground state. This is essential to establish a quantum phase transition from the spin-liquid state (for $x$ = 0 sample) to the spin-glass state with Ca-substitution in $Sr_{1-x}Ca_xNi_2V_2O_8$. Figure 5(a) shows the temperature dependent dc magnetic susceptibility ($\chi_{dc}$) curve for the parent compound $SrNi_2V_2O_8$, measured under an external applied magnetic field of 100 Oe in the ZFC condition. A similar temperature dependent $\chi_{dc}$ curve for the same compound was reported earlier in literatures.[16, 17] With decreasing temperature, the $\chi_{dc}$ ($T$) curve shows a broad maximum around 130 K. With further lowering of temperature, the $\chi_{dc}$ ($T$) curve decreases down to 14 K and then shows an upturn down to lowest measured temperature (2 K). An absence of a peak in the ZFC susceptibility curve (Fig. 5a) confirms that the parent compound does not show any spin glass-like freezing. The observation is consistent with the reported spin-liquid behavior for the parent compound.[16, 17] The observed upturn at lower temperature (below ~ 10 K) is common to most of the quasi-1D magnetic systems and was explained due to the Curie-Weiss contribution arising from isolated magnetic ions present possibly as lattice defects.[16] According to the theoretical prediction,[24] the $\chi_{dc}(T)$ curve for a Haldane system (1D spin-chain system with an integer spin) is proportional to the $\frac{1}{\sqrt{T}}\exp\left(-\Delta/k_BT\right)$ in the low temperature limit ($T \ll \Delta$), where $\Delta$ is the energy gap between singlet ground state and triplet first excited state. We have fitted the low temperature $\chi_{dc}$ ($T$) data with the following equation, by considering contributions from free spins and actual spin contribution of a spin-1 Haldane system.

$$\chi_{dc}(T) = \frac{C}{T-\theta} + \frac{a}{\sqrt{T}}\exp\left(\frac{-\Delta}{k_B T}\right) \tag{1}$$

where, $C$ is the Curie-Weiss constant, $\theta$ is the Curie-Weiss temperature, and $a$ is a proportionality constant. The fitting yields the values of the parameters $C$ = 0.045(3) emu K mol$^{-1}$ Oe$^{-1}$ and $\theta$ = -1.7(2) K. The value of energy gap ($\Delta$) was found to be 26.6(6) K. The values indicated in parentheses are the standard deviations. All these values are in good agreement with the earlier reported values for the same compound SrNi$_2$V$_2$O$_8$.[16] The fraction of isolated Ni$^{2+}$ ions can be estimated from the fitted value of the Curie constant $C$. It is know that $\chi_{mole}$ = ($\mu_0 N_A \mu_{eff}^2$)/3$k_B T$ = $C/T$, therefore, $C = \mu_0 N_A \mu_{eff}^2/3k_B$ where $\mu_0$ is the magnetic permiability at free-space, $N_A$ is the Avogadro's number, and $k_B$ is the Boltzmann's constant. Again, effective paramagnetic moment, $\mu_{eff} = g \sqrt{(S(S+1))} \mu_B$. Now for the present system, S =1, and considering only spin contribution, g =2, and the $\mu_{eff} = 2 \sqrt{(1(1+1))} \mu_B = 2 \sqrt{(2)}\mu_B$, or $(\mu_{eff})^2 = 8\mu_B^2$. Therefore, $C = \mu_0 N_A 8\mu_B^2/3k_B$= 1.0 emu K mol$^{-1}$ Oe$^{-1}$. Now comparing to the experimentally observed value of C, the isolated Ni$^{2+}$ ions is estimated to be 4.5 at. % . The fraction of isolated Ni$^{2+}$ ions is higher than the reported value for the same SrNi$_2$V$_2$O$_8$ compound.[16, 17] Besides, we have estimated the value of the intra-chain exchange constant $J$ from the expression of the susceptibility for a 1D Heisenberg antiferromagnet with $S$ = 1 using the high temperature series expansion as follows,[25]

$$\chi_{spin}(T) = \frac{Ng^2\mu_B^2}{k_B T} \frac{2 + 0.0194x + 0.777x^2}{3 + 4.346x + 3.232x^2 + 5.834x^3} \tag{2}$$

where, $x = |J|/k_B T$ and all other notations carry their usual meanings. The best fit was obtained for the $|J|/k_B$ value of 106.4(2) K (corresponding to $J$ = 9.17 meV) which agrees well with the value 102 K previously reported for the same compound SrNi$_2$V$_2$O$_8$.[16] For the non interacting isotropic spin-1 chains, it is theoretically predicted that the gap energy ($\Delta$) at the AFM zone centre is equal to 0.41$J$ = 43.6 K which is considerably higher than the derived value of 26.6(6) K (obtained from the Eq. (1)) for the present SrNi$_2$V$_2$O$_8$ compound. An inter-chain interaction and/or a single ion anisotropy can play an important role here. Presence of an inter-chain coupling suppresses the Haldane gap, and can induce a 3D long-range magnetic order upon achieving a critical value. Anisotropy is another common perturbation that causes zero-field splitting of the triplet state, and hence a reduction of the effective Haldane gap energy. The powder inelastic neutron scattering study of $SrNi_2V_2O_8$ by Zheludev *et al.*[15] estimated an inter chain interaction ($J_\perp$) of ~ 0.18 meV and a single ion anisotropy ($D$) of ~ -0.56 meV. Now, by considering the single ion anisotropy alone, the expected Haldane gap ($\Delta$) energy for the present compound is estimated below. According to Regnault, *et al.*,[26] for an anisotropy of $D/J$= 0.16, the Haldane gap ($\Delta$) splits into a higher energy singlet state ($\Delta_0$) = 0.62 $J$ and a lower energy doublet state ($\Delta_\pm$)=0.31 $J$, where $\Delta = 1/3(\Delta_0 + 2\Delta_\pm)$. For the studied compound $SrNi_2V_2O_8$, the $|D/J| = |(-0.56 meV)/(106.4 K = 9.17 meV)| = 0.06$. The sign of $D$ causes an alternation of the energy levels. Now, by considering a linear relation between $D/J$ and splitting energy, the value of the lowest energy state $\Delta_0$ for the compound $SrNi_2V_2O_8$ is 0.33 $J$ (= 35.1 K) which is still higher than the observed value of 26.6 K. Therefore, both anisotropy ($D$) and inter-chain interaction ($J_\perp$) are responsible for the observed lower value of energy gap (26.6 K).

The ZFC $\chi_{dc}$ ($T$) curves for the Ca-substituted compounds $Sr_{1-x}Ca_xNi_2V_2O_8$ with $x$ = 0.015, 0.025, 0.04, and 0.05 are shown in Fig. 5(b). Interestingly, a peak in the ZFC $\chi_{dc}$ ($T$) curves for all Ca-substituted compounds at lower temperature ($T$ < 7 K), indicating a magnetic transition, has been observed. This is in contrast to the parent compound having a spin-liquid magnetic ground state without any such transition in the $\chi_{dc}$ ($T$) curve. However, at higher temperatures ($T$ > 20 K), all curves have similar temperature dependence as observed for the parent compound. In fact, the curves overlap on each other above ~ 30 K suggesting no significant change in the strength of the intra-chain interaction $J$ with Ca-substitution along the $c$ axis. The peak temperature for the highest substituted compound $x$ = 0.05 was found to be 6.25 K which is in good agreement with the earlier reported value of 6 K for the same composition from a heat capacity measurement under zero field.[17] In the present study, the peak temperature in the $\chi_{dc}$ ($T$) curve for the lowest Ca-substituted compound *i.e.*, $x$ = 0.015 was found to be 4.5 K. Therefore, it may be concluded that the critical concentration of the Ca substitution for the onset of the transition from the disordered spin-liquid to the magnetically ordered state lies between 0 and 0.015. The field-cooled $\chi_{dc}$ (T) curves for the lowest (x = 0.015) and highest (x = 0.05) Ca-substituted samples have also been shown in Fig. 5(b). A bifurcation between ZFC and FC susceptibility curves below ~ 6.5 and 7.0 K for x = 0.015 and x = 0.05 samples, respectively, has been observed, suggesting a spin-glass transition. It may be noted that for the higher Ca substituted compounds (x ≥ 0.025), the susceptibility values over low temperature range (T < 15 K) are smaller as compared to that for the parent compound. This suggests a lower fraction of the isolated $Ni^{2+}$ ions for the Ca substituted compounds. In literature, an increase in the low temperature susceptibility values has been reported due to increase of the isolated $Ni^{2+}$ ions with increasing Mg concentration in $SrNi_{2-x}Mg_xV_2O_8$ compounds.[27] Therefore, it can be concluded that the isolated $Ni^{2+}$ ions are not responsible for the observed spin-glass behavior as the parent compound (with a higher concentration of $Ni^{2+}$ ions) does not show any spin glass-like freezing. The observed spin-glass behavior possibly arises due to the formation of magnetic polarons (discussed later).

Now, we discuss the nature of the magnetic ground state for the Ca-substituted compounds. For this purpose, a detailed magnetization study as a function of both temperature and external magnetic field has been carried out on the highest Ca-substituted compound x = 0.05. The peak in the $\chi_{dc}$ (T) curves [Fig. 6] shifts towards the lower temperature upon increasing magnetic field. The peak also looses its sharpness and becomes rounded in nature with the increasing applied field. The peak disappears under a magnetic field of ~ 1 kOe, indicating the suppression of the magnetic ordering. The field dependence of the $T_P$, depicted in the inset of Fig. 6, shows a concave type phase boundary in the $T_P$-H plane. Such type of field dependence of $T_p$ indicates a spin-glass type magnetic ordering.[28-30] In the case of a spin-glass system, a spin-freezing transition occurs at a temperature ($T_g$) where spins are aligned in random directions in space. An application of a magnetic field tries to align the spins along the applied field direction, hence, destroys the spin-glass ordering. Therefore, a decrease of $T_g$ occurs with an increasing field strength. A convex type phase boundary in the $T_P$-H plane, in contrary to the observed concave type phase boundary in the present study, was reported for a variety of AFM systems such as the quasi-1D spin-chain compound $BaCo_2V_2O_8$, and the heavy-fermion systems $YbRh_2Si_2$ and $YbNiSi_3$.[31-33]

To understand the nature of magnetic ordering for the present system in detail, we have fitted the $T_p$ vs $H$ curve with the following Eq. (3), generally used for spin-glass systems to define the phase boundary.

$$H(T_g) \propto \left[1 - \frac{T_g(H)}{T_g(0)}\right]^{\beta} \tag{3}$$

where, $T_g(H)$ and $T_g(0)$ are the spin-freezing temperatures under an applied magnetic field and under a zero field, respectively. Here, $T_p$ is considered to be same as $T_g$. The fitted values of the exponent $\beta$, and $T_g(0)$ are found to be 1.47±0.03 and 7.6±0.1, respectively. The value of the exponent $\beta$ is in good agreement with the value $\beta = 3/2$ predicted by the mean field theory for Ising spin-glass systems.[30]

Figure 7 shows the isotherm magnetization curve as a function of applied magnetic field (over all four quadrants), measured at 2 K under zero-field-cooled condition, for the highest Ca-substituted compound ($x = 0.05$). The "S" shape type $M(H)$ curve with a small value of the magnetization (0.045 $\mu_B$/f.u. or 0.023 $\mu_B$/Ni-ion under 7 T) suggests a spin-glass like magnetic ordering for this compound.[34-36] The inset shows an enlarged view of the $M$ vs. $H$ curve over the lower field region. The presence of a hysteresis, with a remanent magnetization of 7 emu/mole and a coercive field of 0.6 kOe, is evident at this temperature [inset of Fig. 7]. The above results of the field dependent magnetization also suggest a spin-glass like magnetic ordering in the studied Ca-substituted compound $Sr_{0.95}Ca_{0.05}Ni_2V_2O_8$.

**C. ac Susceptibility Study:**

In order to confirm the spin-glass nature of the magnetic ordering in the Ca-substituted compounds further, we present here the results of ac susceptibility measurements as a function of both temperature and frequency. The temperature dependence of real part of the ac susceptibility $\chi'_{ac}$ over 1–987 Hz for the compound $Sr_{0.95}Ca_{0.05}Ni_2V_2O_8$ is depicted in Fig. 8(a). The $\chi'_{ac}$ curve shows a peak at a temperature ($T_f$), which shifts towards a higher temperatures from 7 K to 9 K with increasing frequency ($f$) from 1 Hz to 987 Hz. The variation of the peak temperature with frequency is shown in the inset of Fig. 8(a). The shifting of peak position towards higher temperature with increasing frequency is a phenomenon that is observed commonly in spin-glass systems.[37-39] A quantitative measurement of the frequency shift for the susceptibility $\chi'_{ac}(T)$ can be estimated from $s = \Delta T_f/[T_f\Delta\log_{10}f]$, where $\Delta T_f = (T_f1 - T_f2)$ and $\Delta\log_{10}f = [\log_{10}f1 - \log_{10}f2]$ with $f1 = 987$ Hz and $f2 = 1$ Hz. The value of "$s$" is found to be 0.069 which is higher than the values ($10^{-2}$–$10^{-3}$) reported for conventional spin-glass systems[40] The frequency dependence of $T_f$ can be described by the conventional critical ''slowing down'' of the spin dynamics [40, 41] as described by

$$\tau(T_f) = \tau_0 \left(T_f/T_g - 1\right)^{-zv} \qquad (4)$$

where $\tau \sim 1/f$, $T_g$ is the critical temperature for spin-glass ordering at $f \to 0$, $zv$ is a dynamical exponent, and $\tau_0$ is the characteristic time scale for the spin dynamics. The agreement with Eq. (4) is shown in Fig. 8 (b), where $\log_{10}f$ is plotted as a function of $\log_{10}[(T_f/T_g)-1]$. The best fit to the Eq. (4) is obtained by choosing the value of $T_g = 5.2(1)$ K, which minimizes the least-square deviation from a straight-line fit. The values of $\log_{10}(\tau_0) = 4.38(8)$ and $zv = 10.0(3)$ are then extracted from the intercept and slope of the fitted straight line, respectively. The value of $zv$ is in good agreement with the values 9.0–10.0 reported experimentally[41-43] or 7.0–8.0 reported theoretically[44, 45] for Ising spin-glass systems. The obtained characteristic time scale $\tau_0$ ($\sim 10^{-4.38}$ sec) for the present compound is found to be much higher than that in conventional spin-glass systems ($\tau_0 = 10^{-13}$ sec), which is possibly due to a magnetic spin cluster formation, instead of single atomic spins.[39, 46]

In the present study, the detailed $\chi_{dc}(T, H)$, $M(H)$, and $\chi'_{ac}(f, T)$ measurements on the compound $Sr_{1-x}Ca_xNi_2V_2O_8$ with $x = 0.05$, therefore, establish a cluster spin-glass state at lower temperatures ($T \leq 6.5$ K). The observed spin-glass ordering is in contrast to the reported AFM ordering for the same $x = 0.05$ compound.[17] It may be noted that the theoretical study by Sakai et al.[9] predicted a phase transition from the spin-liquid state to a long-range Néel AFM state with the increasing strength of the inter-change interaction $J_\perp$ and/or a reduction in the single ion anisotropy $D$ value. The percentage change of $a$ or $c$ value is less than 0.04% (Ni-Ni distance changes by ~

0.2%); the super-exchange pathway should, therefore, not be affected. A minor cation (Ca) substitution can affect slightly the magnetic anisotropy value $D$ (-0.56 meV for $SrNi_2V_2O_8$)[15], however, may not be sufficient for a phase transition. The observed spin-glass state may be induced by the quenched disorder and/or magnetic polarons. The random insertion of a few percent of different bonds via Ca substitution at the larger Sr site can create magnetic polarons. Upon cooling, glassy correlations may develop due to these polarons which can be magnetized by an application of a uniform magnetic field. This is in contrast to the induced long-range magnetic ordering due to the closing of the Haldane gap under an application of internal or external perturbations. For example, the spin-1 Haldane quasi-1D system $CsNiCl_3$ shows a 3D long-range AFM ordering below $T_N$ =4.4 K due to a sufficiently strong inter-chain interaction.[47] A magnetic field induced magnetic ordering was reported for the iso-structural compound $PbNi_2V_2O_8$.[48] Here, the energy gap is suppressed due to the Zeeman splitting of the triplet state in which the energy of one of the states reduces due to application of the magnetic field and become zero at a critical field. This yields a long-range AFM ordering above the critical field. A non-magnetic impurity doping at the magnetic Ni site also causes a long-range magnetic ordering by liberating $S = 1/2$ end chain degrees of freedom.[14, 19, 21] For the present case, with the Ca-substitution (over the studied concentration range $x \leq 0.05$), no AFM ordering is found; rather a spin-glass state is observed. The polarons may be the origin of the observed spin-glass ordering. The scenario of magnetic ordering (due to closing of the energy gap) induced by a strong inter-chain coupling or the anisotropy is unlikely here. The theoretically predicted long-range Néel AFM ordering may perhaps be stabilized with a further higher concentration of Ca ($x > 0.05$). The present results, therefore, show the necessity of further experimental study on the Ca-substituted compounds with higher concentration ($x > 0.05$) to verify experimentally the theoretically predicted Néel type AFM state.

### IV. SUMMARY AND CONCLUSION:

In summary, both dc and ac magnetic susceptibility studies on the quasi-1D spin-chain systems $Sr_{1-x}Ca_xNi_2V_2O_8$ ($x$ = 0, 0.015, 0.025, 0.04, and 0.05) establish a quantum phase transition from a spin-liquid state (for the parent $x = 0$ sample) to a spin-glass like ground state (for the Ca-substituted samples), induced by a very small amount of Ca-substitution at Sr site. For the parent compound ($x = 0$), dc magnetization study confirms a spin-liquid type disorder magnetic ground state with a finite energy gap of 26.6 K between singlet ground state and triplet excited state. For the Ca-substituted compound ($x = 0.05$), both dc-magnetization and ac-susceptibility studies confirm a spin-glass like ordering. An increase in the spin-glass ordering temperature in the $\chi_{dc}$ ($T$) curves (from 4.5 K for $x = 0.015$ compound to 6.25 K for $x = 0.05$ compound) with Ca-substitution has been found. The critical concentration of Ca for the onset of the quantum phase transition lies between $x = 0$ and $x = 0.015$.

**Table I:** The Rietveld refined lattice constants (*a* and *c*), fractional atomic coordinates, and isotropic thermal parameters ($B_{iso}$) for the samples $Sr_{1-x}Ca_xNi_2V_2O_8$ ($x = 0$, 0.015, 0.025, 0.04, and 0.05).

| | Site | $x = 0$ | $x = 0.015$ | $x = 0.025$ | $x = 0.04$ | $x = 0.05$ |
|---|---|---|---|---|---|---|
| *a* | | 12.1608(1) | 12.1594(2) | 12.1587(1) | 12.1579(1) | 12.1573(2) |
| *c* | | 8.3242(1) | 8.3253(1) | 8.3260(2) | 8.3268(1) | 8.3272(1) |
| **Sr/Ca** | 8a (0, 0, z) | | | | | |
| z/c | | 0 | 0 | 0 | 0 | 0 |
| $B_{iso}$ | | 0.38(1) | 0.39(3) | 0.38(3) | 0.37(2) | 0.38(6) |
| **Ni** | 16b (x, y, z) | | | | | |
| x/a | | 0.3315(5) | 0.3314(3) | 0.3304(7) | 0.3301(5) | 0.3297(5) |
| y/b | | 0.3326(2) | 0.3331(4) | 0.3340(3) | 0.3343(5) | 0.3345(4) |
| z/c | | 0.2261(3) | 0.2250(4) | 0.2249(5) | 0.2242(5) | 0.2236(5) |
| $B_{iso}$ | | 0.35(2) | 0.33(4) | 0.35(3) | 0.35(4) | 0.34(5) |
| **V** | 16b (x, y, z) | | | | | |
| x/a | | 0.2571(2) | 0.2594(3) | 0.2580(3) | 0.2585(4) | 0.2591(7) |
| y/b | | 0.0798(6) | 0.0802(4) | 0.0796(2) | 0.0803(6) | 0.0798(5) |
| z/c | | 0.0966(9) | 0.0964(8) | 0.0959(8) | 0.0957(6) | 0.0959(8) |
| $B_{iso}$ | | 0.29(2) | 0.29(4) | 0.29(6) | 0.27(3) | 0.28(6) |
| **O1** | 16b (x, y, z) | | | | | |
| x/a | | 0.1587(1) | 0.1549(3) | 0.1573(2) | 0.1543(3) | 0.1584(3) |
| y/b | | 0.4964(6) | 0.4975(5) | 0.4980(4) | 0.5004(6) | 0.4988(3) |
| z/c | | 0.0134(2) | 0.0136(4) | 0.0089(3) | 0.0075(6) | 0.0060(5) |
| $B_{iso}$ | | 0.87(3) | 0.89(5) | 0.88(3) | 0.89(2) | 0.89(5) |
| **O2** | 16b (x, y, z) | | | | | |
| x/a | | 0.3470(2) | 0.3380(5) | 0.3418(3) | 0.3387(5) | 0.3392(4) |
| y/b | | 0.6712(5) | 0.6702(5) | 0.6721(8) | 0.6706(2) | 0.6699(3) |
| z/c | | 0.4909(3) | 0.4847(3) | 0.4835(4) | 0.4771(5) | 0.4860(4) |
| $B_{iso}$ | | 0.65(3) | 0.53(8) | 0.62(5) | 0.65(3) | 0.66(6) |
| **O3** | 16b (x, y, z) | | | | | |
| x/a | | 0.1639(2) | 0.1559(3) | 0.1623(6) | 0.1557(5) | 0.1548(3) |
| y/b | | 0.6768(5) | 0.6790(4) | 0.6767(5) | 0.6757(3) | 0.6801(2) |
| z/c | | 0.7259(3) | 0.7218(3) | 0.7209(2) | 0.7134(6) | 0.7224(4) |
| $B_{iso}$ | | 0.97(7) | 0.94(3) | 0.86(8) | 0.76(8) | 0.93(8) |
| **O4** | 16b (x, y, z) | | | | | |
| x/a | | 0.3289(5) | 0.3345(2) | 0.3404(3) | 0.3349(4) | 0.3284(7) |
| y/b | | 0.5026(3) | 0.4995(2) | 0.4964(4) | 0.5030(6) | 0.4969(2) |
| z/c | | 0.2099(6) | 0.2057(5) | 0.1993(5) | 0.2036(7) | 0.2028(5) |
| $B_{iso}$ | | 0.94(8) | 1.01(2) | 0.89(5) | 0.91(4) | 0.94(3) |

**Table II:** The bond lengths for SrNi$_2$V$_2$O$_8$ compound at 300 K.

| Cation | O1 | O2 | O3 | O4 |
|---|---|---|---|---|
| Sr | 2 × 2.776(6)<br>2 × 2.921(10) | 2 × 2.792(4) | 2 × 2.938(11) | 2 × 2.717(9)<br>2 × 3.188(6) |
| Ni | 2.031(7) | 1.967(5)<br>2.188(9) | 2.041(5)<br>2.082(7) | 2.072(4) |
| V | 1.669(6) | 1.730(5) | 1.734(8) | 1.691(10) |

**Table III**: Selected bond angles for SrNi$_2$V$_2$O$_8$ compound at 300 K

| Intra-chain | | Inter-chain | | | | | |
|---|---|---|---|---|---|---|---|
| Ni-O2-Ni | 87.7(5) | Ni-O1-V | 131.2(6) | O1-V-O3 | 118.9(2) | O1-V-O2 | 107.7(4) |
| Ni-O3-Ni | 88.7(2) | Ni-O2-V | 118.7(8) | O2-V-O4 | 102.0(7) | O3-V-O4 | 107.6(6) |
| | | Ni-O3-V | 128.1(5) | | | | |
| | | Ni-O4-V | 126.8(5) | | | | |

# FIGURES

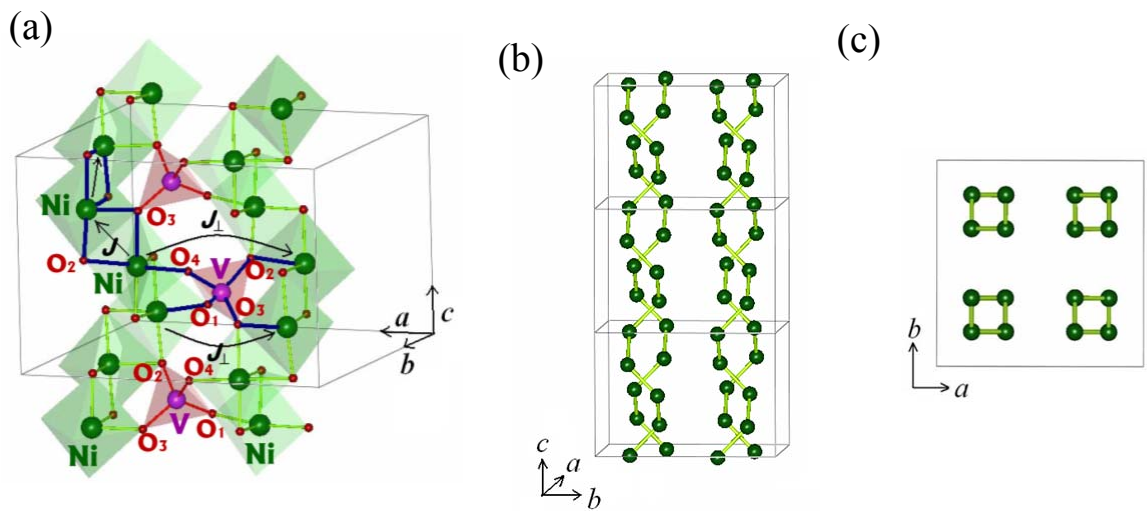

**Fig. 1:** (color online) (a) The crystal structure for the compounds $Sr_{1-x}Ca_xNi_2V_2O_8$ showing the $NiO_6$ octahedral screw chains along the crystallographic $c$ direction and their interlinks by $VO_4$ tetrahedra. Out of four screw chains of $NiO_6$ in a unit cell, only two chains are shown for clarity. Sr/Ca ions are also omitted for simplification. The intra-chain ($J$) and inter-chain ($J_\perp$) superexchange interaction pathways are also shown. (b) The propagation of the four $Ni^{2+}$ screw chains in a unit cell along the $c$ axis. (c) The projection of the screw chains in the $ab$ plane. The dimension of the unit cells is shown by thin grey lines.

Page 19 of 26

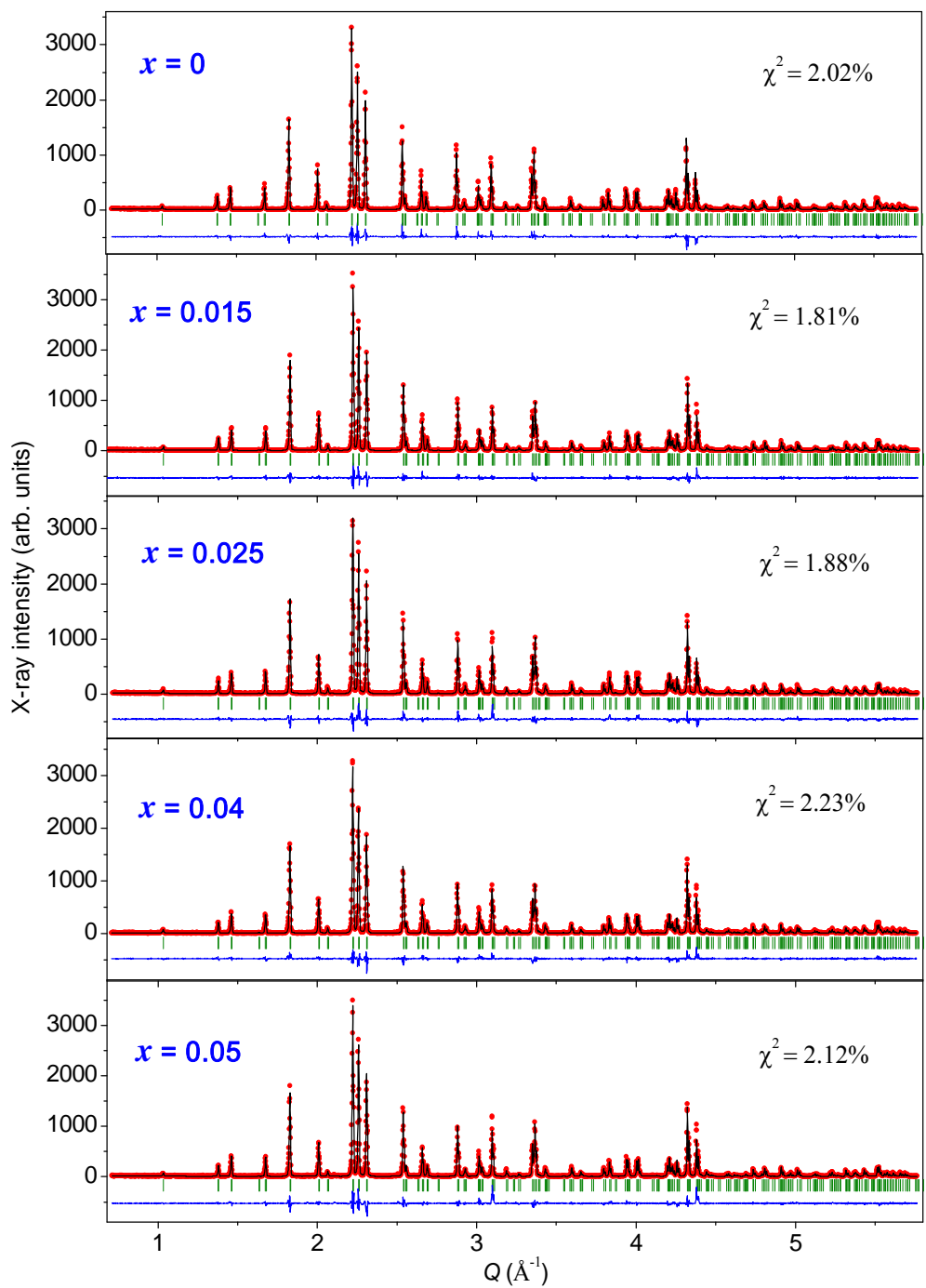

**Fig. 2:** (color online) Observed (solid circles) and calculated (solid line) XRD patterns for $Sr_{1-x}Ca_xNi_2V_2O_8$ ($x$ = 0, 0.015, 0.025, 0.04, and 0.05) compounds at 300 K. Solid line at the bottom of each panel shows the difference between observed and calculated patterns. Vertical lines show the positions of Bragg peaks.

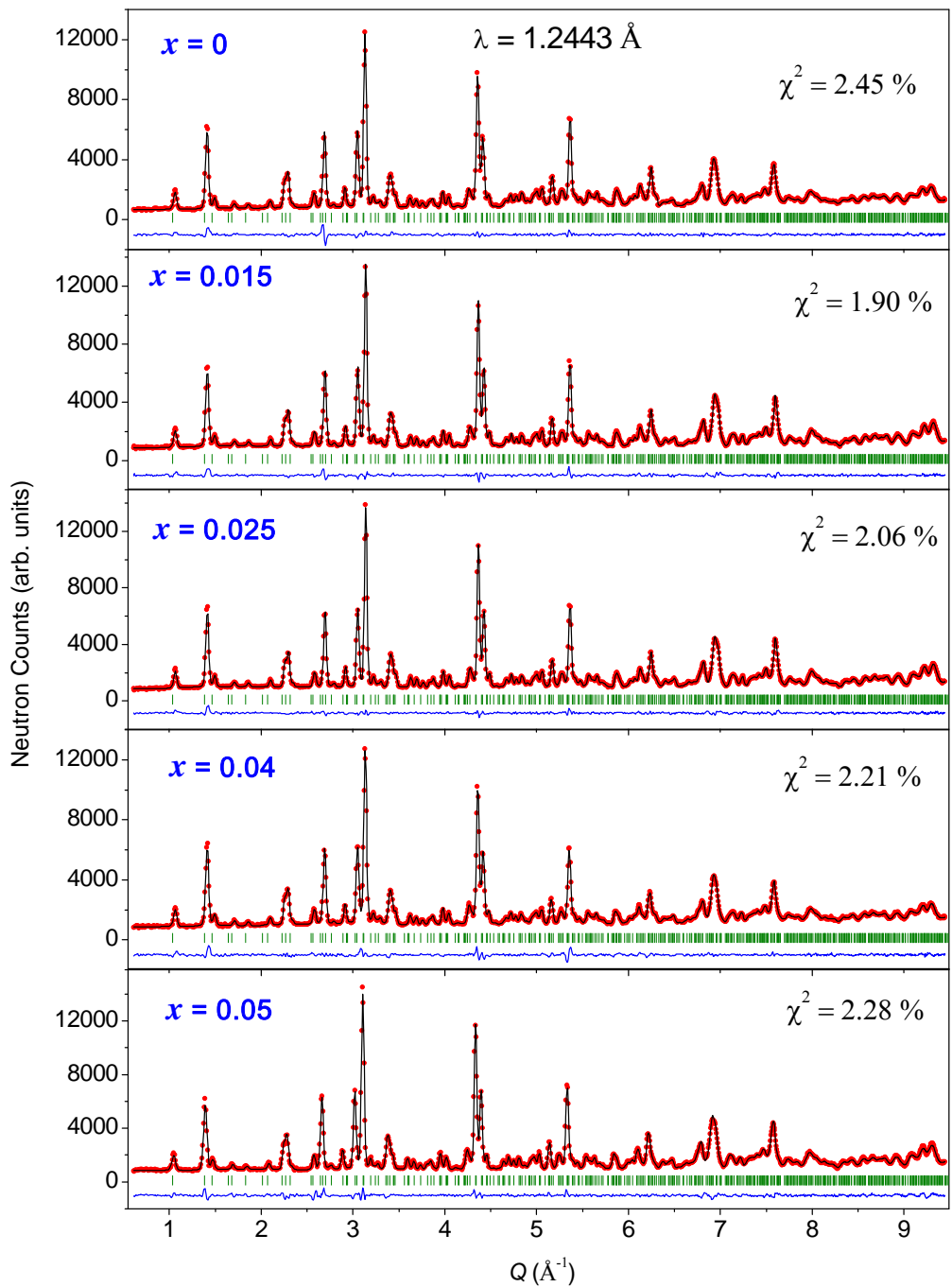

**Fig. 3:** (color online) Observed (solid circles) and calculated (solid line) neutron powder diffraction patterns for $Sr_{1-x}Ca_xNi_2V_2O_8$ ($x$ = 0, 0.015, 0.025, 0.04, and 0.05) compounds at 300 K. Solid line at the bottom of each panel shows the difference between observed and calculated patterns. Vertical lines show the positions of Bragg peaks.

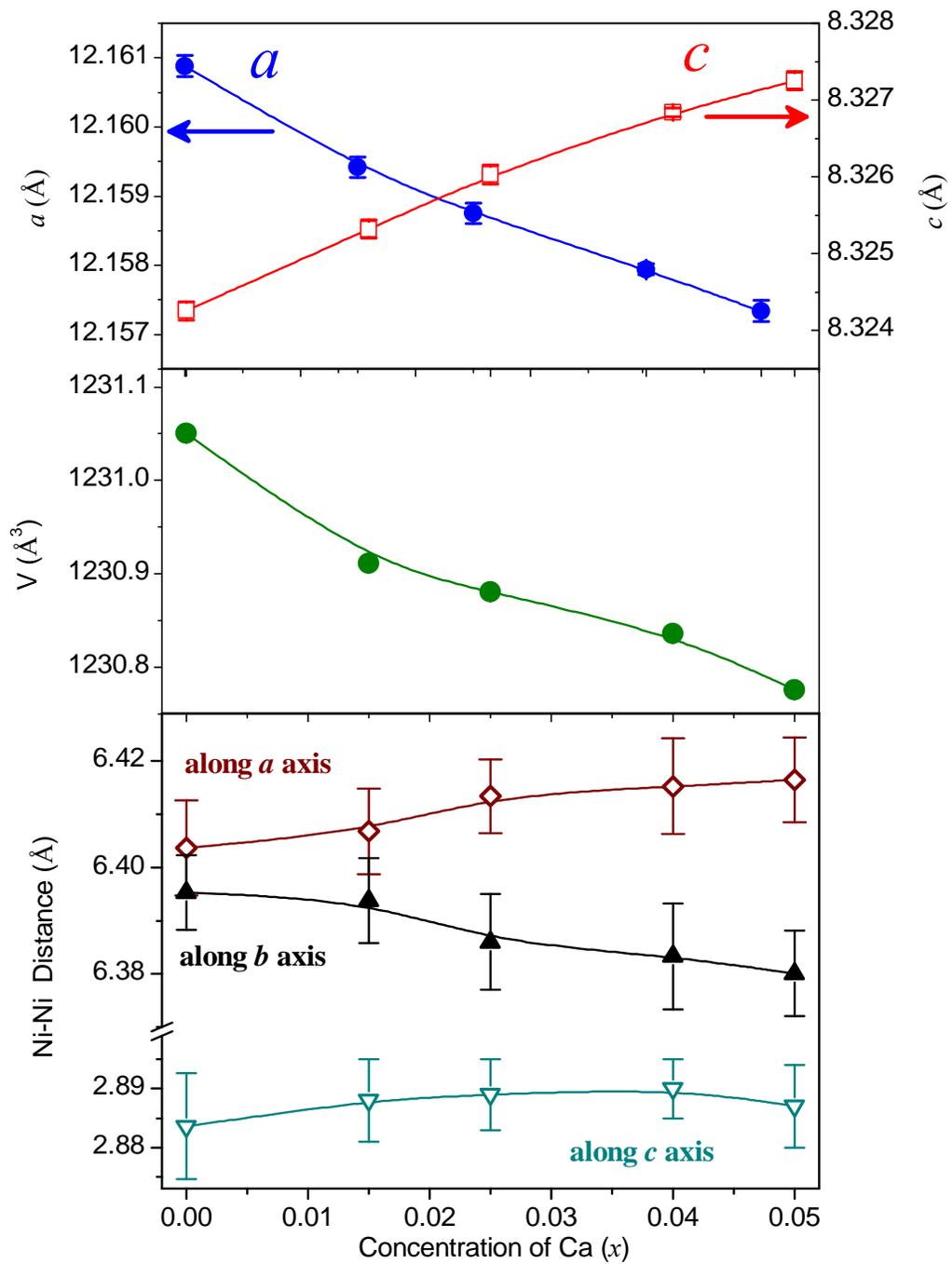

**Fig. 4:** (color online) The variation of the lattice constants (*a* and *c*), unit cell volume (*V*), and Ni-Ni direct distances with the Ca-concentration (*x*).

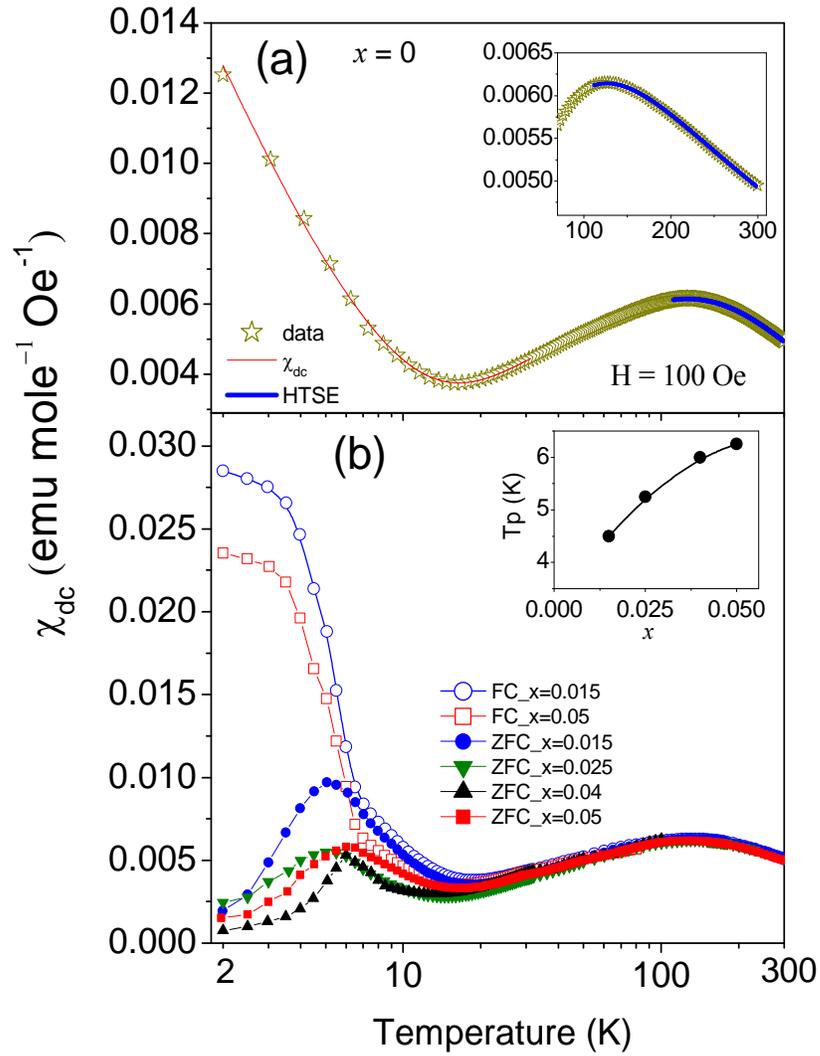

**Fig. 5:** (color online) (a) Temperature dependent dc susceptibility ($\chi_{dc}$) curve (measured in ZFC condition) for parent compound SrNi$_2$V$_2$O$_8$. The solid lines represent the theoretical fitting to the observed data with Eq. (1) (thin line below 25 K) and Eq. (2) (thick line above 100 K). Inset shows an enlarge view of the $\chi_{dc}(T)$ curve over 80-300 K to show the HTSE fit. (b) Temperature dependent ZFC $\chi_{dc}$ curves for all Ca-substituted compounds Sr$_{1-x}$Ca$_x$Ni$_2$V$_2$O$_8$ with $x$ = 0.015, 0.025, 0.04, and 0.05. The FC $\chi_{dc}$ curves are also shown for the lowest ($x$ = 0.015) and the highest ($x$ = 0.05) Ca-substituted samples. Inset shows the variation of the peak temperature ($T_P$) as a function of Ca concentration ($x$).

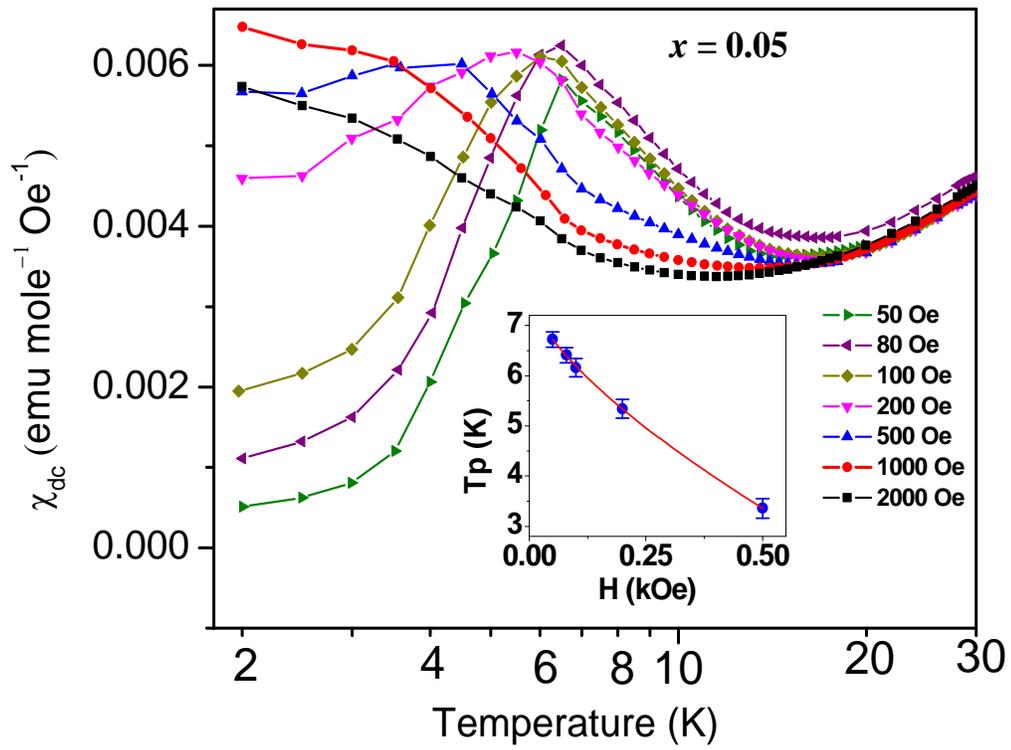

**Fig. 6:** (color online) Temperature dependent dc susceptibility ($\chi_{dc}$) curves for the highest Ca-substituted compound $Sr_{0.95}Ca_{0.05}Ni_2V_2O_8$ measured under different external magnetic fields. Inset shows the dependence of the peak temperature ($T_P$) on magnetic field ($H$). The solid line is the fitted curve according to the Eq. (3).

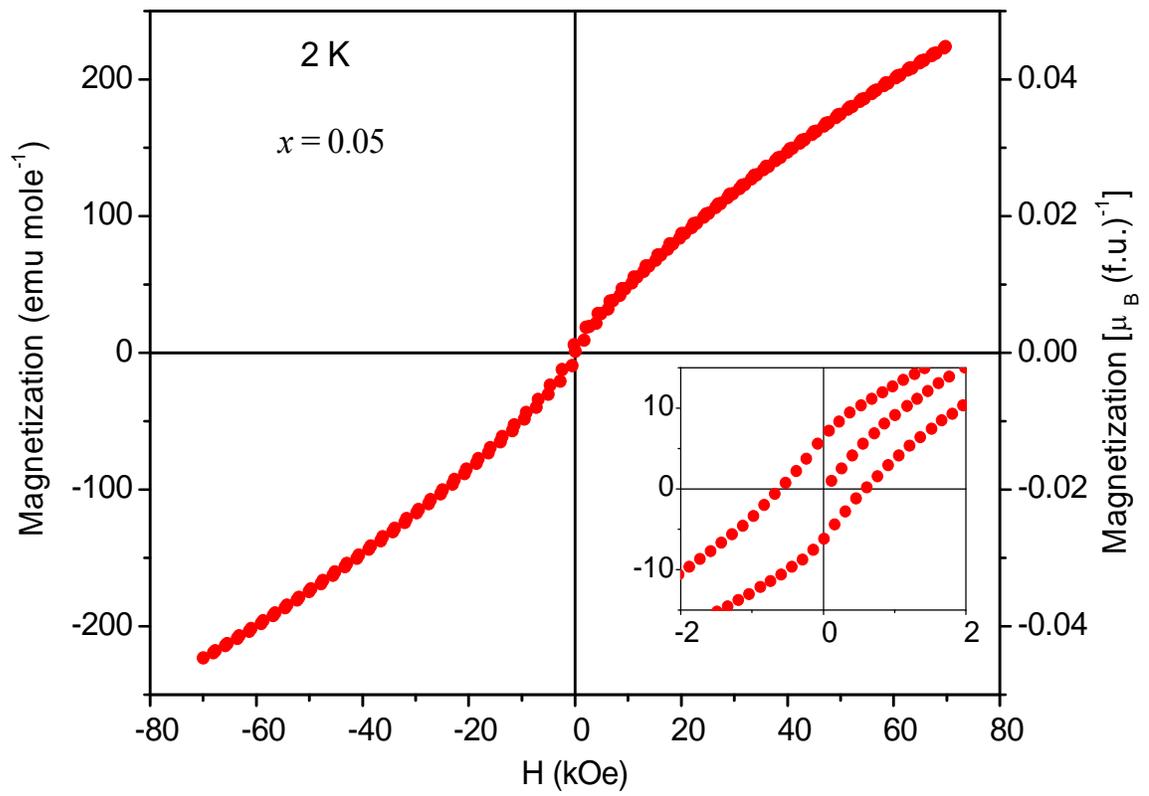

**Fig. 7:** (color online) The *M* vs. *H* curve over four quadrants, measured at 2 K, for the Sr$_{0.95}$Ca$_{0.05}$Ni$_2$V$_2$O$_8$ compound. Inset shows an enlarge view of the *M* vs. *H* curve over the low field region.

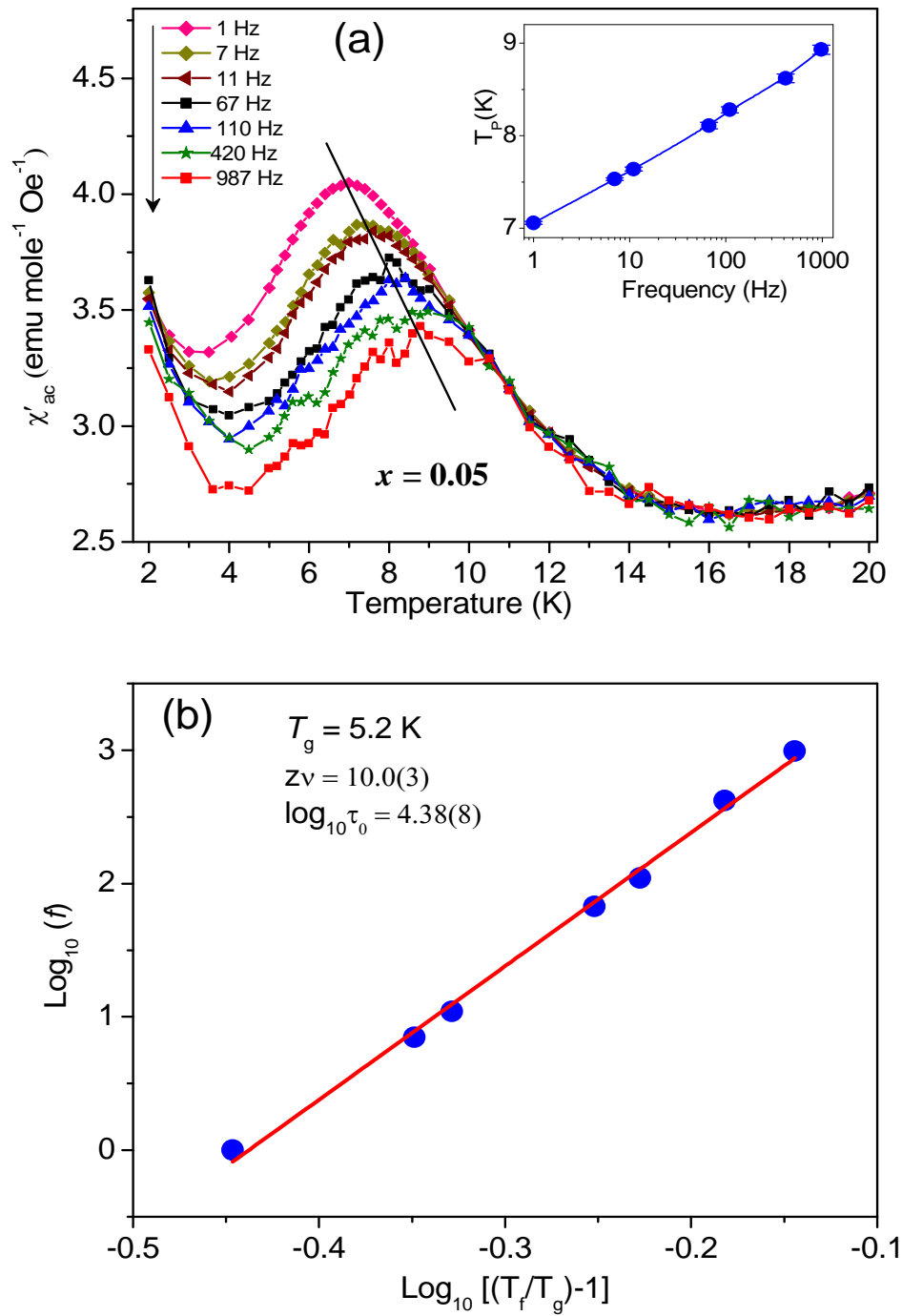

**Fig. 8:** (color online) (a) Temperature dependent real part of the ac susceptibility ($\chi'_{ac}$) curves for the highest Ca-substituted compound $Sr_{0.95}Ca_{0.05}Ni_2V_2O_8$ measures at 1, 7, 11, 67, 110, 420, and 987 Hz with an ac field of 1 Oe. Inset shows the variation of the peak temperature with frequency. (b) $\log_{10}(f)$ vs $\log_{10}[(T_f/T_g)-1]$ plot for the $Sr_{0.95}Ca_{0.05}Ni_2V_2O_8$ compound, demonstrating the agreement with Eq. (4). The solid line is a best fit to the data with the parameters shown in figure.